# *Evolution of article processing charges in medicine: disciplinary position, access gaps and future prospects*

**Juan José Prieto-Gutiérrez.** jjpg@ucm.es
https://orcid.org/0000-0002-1730-8621
**Prof. Manuel Blázquez-Ochando** manublaz@ucm.es
https://orcid.org/0000-0002-4108-7531
**Prof. María Antonia Ovalle-Perandones** maovalle@ucm.es
https://orcid.org/0000-0002-6149-4724

***Complutense University of Madrid***

**Preprint version. This manuscript has not been peer reviewed or previously published. It is made publicly available to facilitate scholarly discussion and may be revised prior to submission for formal publication.**

## Abstract.

Medicine occupies a singular position in the open access economy: it is the field with the highest number of OA journals with APCs in DOAJ, the second largest by global APC expenditure with high rates of APC adoption among researchers in the Global South. This article analyses APC trajectories in medicine from a comparative disciplinary perspective, drawing together evidence from multiple studies to position medicine in relation to other scientific fields over the period 2009–2023. It documents price gaps between medical and biomedical subcategories, the particularly acute affordability challenges associated with publishing in global health journals and the structural asymmetry between the public urgency of medical knowledge and the cost of its dissemination. The article concludes with a reflection on alternative models and the conditions required for an equitable and sustainable open access transition in medicine.

**Keywords:** article processing charges; open access; medical journals; scholarly publishing; global health equity;

## 1. Introduction

The establishment of an explicit fee for the publication of an open-access article marks one of the defining moments in the contemporary economy of scientific knowledge. The field of medical and health journals provides the earliest figures, modest by today's standards but significant in terms of market signals. It was in 2003 that the *Journal of Translational Medicine* explicitly formalised the charging of fees for open-access publication, with an APC of $525 (Marincola, 2003).

This is not a contemporary phenomenon, but rather the culmination of a historical evolution dating back to page charges (for example, the American Physical Society charged $2.00 per page at the time), introduced in 1930 by US scientific publishers as a measure of financial survival for scientific societies during the Great Depression. (Barton, 1963).

Article processing charges (APCs) do not carry the same weight across all scientific disciplines. Their economic impact, their uptake by researchers and the barriers they create to publication vary substantially depending on the field, the publishing traditions of each area and the available sources of funding. Within this varied disciplinary context, medicine occupies a particularly significant position: neither in terms of cost nor volume can it be equated without qualification with the academic system as a whole, and its particular characteristics—the scale of biomedical output, the massive involvement of public and private funding agencies, the social demand for open access to clinical knowledge, and the profound inequalities between regions of the world in terms of ability to pay—make it the field where the tensions of the APC model are most clearly expressed.

The aim of this study is to analyse the position of medicine within the disciplinary landscape of APCs, to document its evolution over time based on available empirical data, and to examine the implications of that position for equity in the production and dissemination of medical knowledge. The analysis is based on a set of studies with verified primary data covering the period 2009–2023, which, for the first time, enable a longitudinal and comparative view of medicine within the APC ecosystem to be constructed.

## 2. Medicine in the disciplinary landscape of apcs

To understand the position of medicine within the APC system, it is necessary to situate it in relation to other disciplines, in terms of both price and volume, as well as the adoption of the model. The available data paint a consistent picture across several years and different geographical contexts.

In the early days, BioMed Central saw growth in the number of OA journals and APCs, as the number of manuscripts submitted to that publisher rose from approximately 200 in 2000 to 1,200 in 2006, —a 500% increase in six years—, a direct reflection of the progressive legitimisation of the OA model with APCs within the biomedical scientific community (Cockerill, 2006).

As for the average price, Klebel and Ross-Hellauer (2023), based on an analysis of OA publications between 2009 and 2016, place medicine, with an average APC of $1,813, in third place on a scale of 19 disciplines led by biology ($2,118) and chemistry ($1,824). This position is stable and consistent with that documented for 2015 by Solomon and Björk (2016), who, based on data on actual payments in European and North American sources, calculated that the average APC in clinical medicine in hybrid journals was $3,000.33 (n = 626) and in biomedical research of $2,996.56 (n = 1,377), whilst in pure open-access journals these figures fell to $1,870.32 and $1,952.02 respectively.

The field of psychiatry and psychology, situated at the intersection of clinical medicine and the social sciences, recorded an average APC of $2,955.87 for the hybrid model and $1,647.01 for the pure OA model in the same year—above the overall average, but below that of the clinical specialities with the highest impact.

The disciplinary structure of the market also exhibits an analytical pattern that the literature on APCs does not always make explicit: the gap between the price of the pure OA model and that of the hybrid model is particularly pronounced in medicine. Björk and Solomon (2014) documented that, for Elsevier's hybrid journals, the Biomedicine category had an average APC of $2,551, ranging from $1,100 to $5,000. This pattern is consistent with the distribution of journals by discipline analysed by Ghane et al. (2020) for the natural sciences in the DOAJ: of the 1,881 journals studied across 13 scientific subclasses, only 12.9% (243 titles) used the APC model, whilst 87.1% operated without it. In the subclasses with the greatest medical presence—physiology, microbiology, general biology—the proportion of journals with APCs was above average, whilst in mathematics (37 out of 556, 6.7%) and zoology (14 out of 118, 11.9%) it was notably lower. This suggests that the APC model is not universal across the sciences: it is selective, and medicine is one of its epicentres.

Morrison (2021), drawing on the universe of journals indexed in the DOAJ directory for 2019, confirms that the 'Medicine' category is the largest in that directory: 3,347 journals, 291,940 articles published that year, and a weighted average APC per article of $1,753. No other DOAJ category comes close to this output. The proportion of medical journals charging APCs—1,626 out of 3,347, or 48.6%—is the highest of all the major categories in the directory and significantly exceeds the overall DOAJ average (29.0%). Medicine is, in this sense, the field where the pay-per-publication model has the greatest structural penetration within the universe of OA journals. Furthermore, researchers in medicine and health display the least hostile attitude towards the APC system compared to other fields such as the humanities or the social sciences (Segado-Boj et al., 2022).

From the perspective of total aggregate expenditure, Butler et al. (2023), in their analysis of 505,555 publications from five major publishers during the period 2015–2018, quantified expenditure on APCs in the Medical and Health Sciences at $423.1 million, the second-highest figure by field, behind only the Natural Sciences ($450.5 million). The difference between the two fields is just 6.4%, placing medicine on a par with the basic sciences in practical terms, despite having a lower overall output.

The APC review up to 2023 concludes that, for example, in biomedicine, positions are primarily funded through project grants, institutional funds and 'read-and-publish' agreements, with a smaller proportion coming from direct self-funding (Borrego, 2023).

## 3. Geographical and equity gaps: publishing medicine from the Global South

The position of medicine in the ranking of APCs by discipline takes on a different political dimension when analysed from the perspective of researchers in the Global South. In 2023, a survey conducted by the Global Research Institute of Paris, cited by Gallardo et al. (2024), documented that Medical and Health Sciences had the highest adoption rates of the APC model of all fields in the four countries analysed: 71.2% in Argentina, 85.1% in Mexico, 87.8% in Brazil and 90.8% in South Africa. This adoption rate, which in all cases exceeds that of disciplines such as Engineering and Technology or Basic and Natural Sciences, does not reflect greater availability of financial resources, but rather the pressure of open access mandates from international biomedical funding agencies that co-fund a significant proportion of medical research in these countries.

The tension between this institutional pressure and actual ability to pay is particularly acute in the field of global health. Edem et al. (2021) documented the article processing charges (APCs) of 13 international global health journals, with figures ranging from €975 for the *Journal of Epidemiology and Global Health* to $5,000 for *The Lancet Global Health*. Nabyonga-Orem (2020), meanwhile, reported that the average monthly income of specialist doctors in nine African countries was $3,480, with a minimum of $449 in Ethiopia. Cross-referencing both datasets reveals an asymmetry of moral significance: for an Ethiopian medical researcher, publishing in the highest-impact global health journal equates to more than eleven months' gross salary. For the average African specialist in the sample, that same APC represents 1.4 months' salary. This ratio illustrates an especially acute affordability problem in medical publishing and places medicine before a deontological paradox of the first order: the knowledge most urgently needed by the most vulnerable populations is also the most expensive to publish for the researchers working with those populations.

The figures provided by Klebel and Ross-Hellauer (2023) on the relationship between average article processing charges (APCs) paid and GDP per capita add a systemic dimension to this anomaly. Researchers from countries with low to medium GDP—such as India and Brazil—do not pay the same APCs as those from Switzerland or the United States because they publish predominantly in journals with lower impact factors or benefit from partial exemptions. The self-selection effect this generates is not neutral: it shifts medical research output from the Global South towards journals with lower international visibility, perpetuating an epistemic hierarchy in which clinical knowledge generated in contexts of high disease burden remains systematically less visible than that produced in high-income settings. Waiver mechanisms—the route that should, in principle, compensate for this asymmetry—do not always resolve the structural problem. Evidence from the biomedical field shows that partial discounts may remain insufficient: even a 50% reduction in APCs can leave publication costs beyond the reach of researchers with limited funding (Rouhi et al., 2022). Some major publishers have sought to address this limitation by offering full APC waivers to authors from eligible countries, typically using criteria based on Research4Life or World Bank classifications. Although these policies can remove the immediate payment barrier for eligible authors, they do not fully resolve inequalities in access to publication, since eligibility criteria may exclude researchers who fall outside the designated countries or funding categories. Moreover, the broader publishing environment has also seen the emergence of predatory journals using misleading or false waiver offers, further complicating the choices available to researchers with limited resources.

Indeed, this system may discourage international collaboration, as the presence of a co-author from a high-income country tends to automatically invalidate the exemption. (Erondu et al., 2020).

Taking all this into account, it could be summarised that the position of medicine and its subcategories reveals several patterns: on the one hand, the price in medicine is consistent and high across all periods and sources, higher than in engineering and the social sciences, and close to biology and chemistry in absolute terms, but with the particularity that its higher volume of production makes it the second-highest global expenditure on APCs; on the other hand, the gap between pure and hybrid OA in clinical medicine is one of the widest across the entire system, indicating that medicine is a field where the hybrid model generates revenue based on existing institutional subscriptions, and the extensive information from disaggregated data in the field of medicine is viewed with great interest within the paid OA market.

Taken together, these findings reveal several distinctive patterns in the position of medicine and its subcategories within the APC landscape. APC levels in medicine remain consistently high across the periods and sources analysed: they are generally higher than those observed in engineering and the social sciences and close to those of biology and chemistry in absolute terms. However, medicine combines these relatively high prices with a particularly large volume of scientific output, making Medical and Health Sciences the second-highest field in terms of aggregate APC expenditure. A further distinctive feature is the substantial price gap between pure and hybrid open access in clinical medicine, with hybrid journals consistently displaying higher APCs. Overall, the disaggregated evidence suggests that medicine occupies a particularly significant position within the paid open access market, not because it systematically records the highest APCs, but because high publication volumes, widespread adoption of the APC model and relatively high publication charges converge within the same disciplinary field.

## 4. Conclusions and outlook

The data examined allow us to draw a clear conclusion: medicine is both the field with the highest uptake of the APC model and the one with the greatest internal debate regarding the principles that underpinned that model. Open access in medicine was promoted on the grounds that the democratisation of clinical knowledge would benefit healthcare systems worldwide. Two decades later, this ideal coexists with a system in which publishing in the field's highest-impact journals carries a cost equivalent to several months' salary for a research physician in the countries where such research and its dissemination are most urgently needed. The search for a balance between the economic sustainability of the publishing system, affordability for authors and universal accessibility for readers must be an unavoidable priority for the medical academic community, funding agencies and the researchers themselves.
Looking ahead, several trends point towards possible transformations of the model over the next decade. The first is the consolidation of transformative agreements or 'read and publish' contracts at national level, already in place in Germany, the Netherlands, the United Kingdom and several Nordic countries.

These agreements shift the payment burden from individual researchers to library institutions, negotiating a single price for read access plus the open access publication of articles by the institution's researchers. Although they reduce the individual payment barrier, they do not eliminate the underlying structural problem: pricing power remains with the major publishers, and several studies show that total costs per article under these agreements remain in the range of $3,500–$4,600, well above the actual costs of publishing.
In second place, the Global South, APCs are high relative to available income and budgets; they reduce the presence of local authors in global health and clinical specialty journals, drive research towards closed models or less visible platforms, and force reliance on waivers, co-authors from the North, or regional journals. The most robust proposals advocate for models without APCs, automatic exemptions linked to a country's income, and specific funding for health-related publications to avoid widening global gaps in medical research.

The third trend is the expansion of the diamond open access model, funded through public or institutional grants, with no charges for either readers or authors. In the medical field, this model has strong historical precedents—the PubMed Central platform in the United States, SciELO in Latin America—but faces the challenge of visibility: diamond-model journals are not usually indexed in the same reference catalogues as those of major commercial publishers, which limits their perceived bibliometric impact and, consequently, their appeal to researchers subject to evaluation systems that reward publication in high-impact-factor journals.

The fourth trend, of a more structural nature, is the growing pressure on scientific evaluation systems to stop using journal impact factor as a proxy for quality. The San Francisco Declaration on Research Assessment and the Leiden Declaration have laid the theoretical foundations for this reform; the European Commission's Research Assessment Reform Agenda (CoARA) and the adoption of qualitative metrics by several national evaluation agencies represent concrete steps in this direction. If academic recognition systems cease to specifically reward publication in high-APC journals, the pressure felt by medical researchers to publish in such outlets — which currently justifies the payment of APCs that can exceed $5,000 — would be substantially reduced.

In this context, library institutions play a strategic role that goes beyond the management of subscription contracts or funds for the payment of article processing charges (APCs). The academic library is the most appropriate institutional actor to coordinate publication policies that balance the international visibility of medical output with financial sustainability and equitable access: by negotiating transformative agreements, funding diamond open access publishing infrastructures, advising researchers on alternatives to APCs, and promoting deposit in open access institutional repositories. Preventing the financial barriers to OA publishing from becoming obstacles to the dissemination of medical knowledge is, ultimately, a shared responsibility between the researchers themselves, academic institutions, funding agencies and public administrations which, in many countries, fund both biomedical research and the health systems that should benefit from it.

The available evidence suggests that no single actor can resolve the underlying contradiction of a publishing system that seeks to make medical knowledge universally accessible while making publication in many of its most visible outlets dependent on the ability to pay. Reforming the APC model in medicine is not a technical issue of contract management; it is often a political issue of access to knowledge in many countries, such as those in the Global South. Resolving it requires that sustainability, affordability and accessibility be treated as simultaneous and equally indispensable objectives, rather than as variables that offset one another in bilateral negotiations between publishers and institutions.

**Suggested citation:**

Prieto-Gutiérrez, J. J., Blázquez-Ochando, M., & Ovalle-Perandones, M. A. (2026). *Evolution of article processing charges in medicine: Disciplinary position, access gaps and future prospects* [Preprint]. arXiv. DOI: